\documentclass[a4paper,12pt]{article}
\usepackage{epsfig}\parskip 5pt plus 1pt
\usepackage{amsmath}
\usepackage{amssymb}
\usepackage{amsfonts}
\usepackage{fleqn} 
\usepackage{mathrsfs}
\usepackage{cite}

\newcommand{\capdef}{}
\newcommand{\mycaption}[2][\capdef]{\renewcommand{\capdef}{#2}%
       \caption[#1]{{\footnotesize #2}}}

\def\I{\mathrm{i}}

\def\beq{\begin{equation}}
\def\eeq{\end{equation}}
\def\bea{\begin{eqnarray}}
\def\eea{\end{eqnarray}}

\def\<{\left\langle}
\def\>{\right\rangle}

\begin{document}

\bibliographystyle{OurBibTeX}

\begin{titlepage}

\vspace*{-15mm}
\begin{flushright}
CERN-PH-TH/2007-036\\
FTUAM 07-03\\ 
IFT-UAM/CSIC 07-08\\
SHEP/07-05\\
hep-ph/0702286
\end{flushright}
\vspace*{1mm}

\renewcommand{\thefootnote}{\it\alph{footnote}}

\begin{center}
{\bf\Large Neutrino Mixing Sum Rules \\[2mm] and Oscillation Experiments} \\[10mm]
{\bf
S.~Antusch\footnote{E-mail: \texttt{antusch@delta.ft.uam.es}},
P.~Huber\footnote{E-mail: \texttt{phuber@physics.wisc.edu}},
S.~F.~King\footnote{E-mail: \texttt{sfk@hep.phys.soton.ac.uk}}
and T.~Schwetz\footnote{E-mail: \texttt{schwetz@cern.ch}}}
\\[5mm]
{\small\it
$^a$ Departamento de Fisica Te\'orica C-XI and Instituto
del Fisica Te\'orica C-XVI,\\
Universidad Aut\'onoma de Madrid, Cantoblanco, E-28049 Madrid, Spain\\
$^b$ Department of Physics, University of Wisconsin,\\
1150 University Avenue, Madison, WI 53706, USA\\
$^c$ School of Physics and Astronomy, University of Southampton,\\
Southampton, SO17 1BJ, UK\\
$^d$ CERN, Theory Division, Physics Department, CH-1211 Geneva 23, Switzerland}
\end{center}
\vspace*{0.40cm}

%%%%%%%%%%%%%%%%%%%%%%%%%%%%%%%%%%%%%%%%%%%%%%%%%%%%%%%%%%%%%%%%%
\begin{abstract}

\noindent
The neutrino mixing sum rule $\theta_{12} - \theta_{13}\cos(\delta)
\approx \theta^\nu_{12}$ provides a possibility to explore the
structure of the neutrino mass matrix in the presence of charged
lepton corrections, since it relates the 1-2 mixing angle from the
neutrino mass matrix, $\theta_{12}^\nu$, to observable parameters of
the PMNS mixing matrix. The neutrino mixing sum rule holds if the
charged lepton mixing angles are CKM-like, i.e., small and dominated
by a 1-2 mixing, and for small 1-3 mixing in the neutrino mass
matrix. These conditions hold in a wide class of well motivated
flavour models.  We apply this sum rule to present oscillation data,
and we investigate the prospects of future neutrino facilities for
exploring the sum rule by simulating various setups for long-baseline
reactor and accelerator experiments.  As explicit examples, we use the
sum rule to test the hypotheses of tri-bimaximal and bimaximal
neutrino mixing, where $\theta^\nu_{12}$ is predicted by
$\sin^2(\theta^\nu_{12}) = 1/3$ or $1/2$, respectively, although the
neutrino mixing sum rule can be used to test any prediction for
$\theta^\nu_{12}$.
\end{abstract}

\end{titlepage}
\newpage
\renewcommand{\thefootnote}{\arabic{footnote}}
\setcounter{footnote}{0}

%%%%%%%%%%%%%%%%%%%%%%%%%%%%%%%%%%%%%%%%%%%%%%%%%%%%%%%%%%%%%%%%%%%%%%%
\section{Introduction} 
%%%%%%%%%%%%%%%%%%%%%%%%%%%%%%%%%%%%%%%%%%%%%%%%%%%%%%%%%%%%%%%%%%%%%%%

Many attractive classes of models of fermion
masses and mixing lead to interesting predictions for the neutrino
mass matrix $m_\nu$, such as for instance tri-bimaximal or bimaximal
mixing. However, the experimentally accessible quantity is the product
$U_{\mathrm{PMNS}} = V_{e_\mathrm{L}} V^\dagger_{\nu_\mathrm{L}}$. It
includes the neutrino mixing matrix $V_{\nu_\mathrm{L}}$, which
diagonalizes $m_\nu$, and the charged lepton mixing matrix
$V_{e_\mathrm{L}}$, which diagonalizes the charged lepton mass matrix
$M_e$. Often, the essential predictions of flavour models are hidden
due to the presence of the charged lepton corrections. 

Remarkably, in many cases it can be shown that a combination of the
measurable PMNS parameters $\theta_{12}$, $\theta_{13}$ and $\delta$
sums up to the theoretically predicted value for the 1-2 mixing of the
neutrino mass matrix.  For example, in an SO(3) family symmetry model
based on the see-saw mechanism with sequential dominance, predicting
tri-bimaximal neutrino mixing via vacuum alignment, the neutrino
mixing sum rule which is the subject of this paper was first 
observed in the first paper of Ref.~\cite{sumrule}.  In the second
and third papers of Ref.~\cite{sumrule}, it was shown that this neutrino mixing
sum rule is not limited to one particular model, but applies 
under very general assumptions, to be specified
below.  For general 1-2 mixing $\theta^\nu_{12}$ of the neutrino mass
matrix, the neutrino mixing sum rule of interest here was given as
\cite{sumrule}:
\begin{eqnarray}
\label{eq:sumrule}
\theta_{12} - \theta_{13}\cos (\delta) &\approx&  \theta^\nu_{12} \; ,
\end{eqnarray}
in the standard PDG parameterization of the PMNS matrix
\cite{PDG}. The specific neutrino mixing sum rules for
tri-bimaximal~\cite{tribi} and bimaximal~\cite{bimax} neutrino mixing
are obtained by replacing $\theta^\nu_{12}$ by its predictions
$\arcsin (\tfrac{1}{\sqrt{3}})\approx 35.3^\circ$ and
$\tfrac{\pi}{4}=45^\circ$, respectively.

Let us note at this point that corrections to neutrino mixing angles
from the charged lepton sector have been addressed in various previous
studies. Since some of them are sometimes also referred to as ``sum
rules'', we would like to comment on the differences to the neutrino
mixing sum rule of Eq.~(\ref{eq:sumrule}) \cite{sumrule}.  For
instance, in many works, it has been noticed that charged lepton
corrections can make bimaximal neutrino mixing compatible with
experimental data \cite{refs1,refs2,refs7,refs3}.  However, in most studies
in the literature (e.g.\ in Ref.~\cite{refs1}), CP phases have been
ignored.  In other works, where complex mixing matrices were
considered, the connection to the experimentally measurable Dirac CP
phase $\delta$ has not been identified~\cite{refs2}. 
For instance, in Ref.~\cite{refs7} in Eq.~(25), the
correction is related to a phase $\phi$, which is not identical to
$\delta$. The introduction of the measurable quantity $J_{CP}$ in this
equation leads to a sign ambiguity in their ``sum rule''. 
We would like to remark that various parameterizations of the PMNS matrix are customary, 
and that it is important to specify unequivocally which convention is used. 
Assuming standard PDG parameterization of the PMNS matrix \cite{PDG} 
where not stated otherwise, the relations of Ref.~\cite{refs3}   
for bimaximal neutrino mixing are physically inequivalent to the sumrule in Eq.~(\ref{eq:sumrule}) \cite{sumrule}. 
Equations for corrections to tri-bimaximal neutrino mixing as well as for
general $\theta^\nu_{12}$ have been considered \cite{refs5,refs6},
however the connection to the measurable CP phases has not been
established.  Finally, CKM-like corrections to neutrino mixing angles
have been considered, and called ``sum rules'', in Ref.~\cite{refs4},
however CP phases have been ignored.  
In the following, we will simply refer to the formula of Eq.~(\ref{eq:sumrule}) \cite{sumrule} as the sum rule.

In this paper, after deriving
 the sum rule \cite{sumrule}, we investigate how well the
  combination of parameters on the left-hand of Eq.~(\ref{eq:sumrule}) can be determined
  in present and future neutrino oscillation facilities, and then
  compare to the predictions for the right-hand side
  coming from bi-maximal and tri-bimaximal neutrino mixing. 
Such a study is interesting since the  sum rule of Eq.~(\ref{eq:sumrule})
\cite{sumrule} provides a means of exploring the structure of the
neutrino mass matrix in the presence of charged lepton corrections,
and enables whole classes of models of neutrino masses and mixings
to be tested. However,
exploring the  sum rule requires to measure the currently undetermined
mixing angle $\theta_{13}$ as well as the CP violating phase $\delta$,
which is experimentally challenging, as we shall discuss.

The outline of the remainder of the
paper is as follows. In Sec.~\ref{sec:sumrule}
we present for the first time
a parameterization-independent derivation of a family
of neutrino mixing sum rules, and subsequently
show that one of them leads to the
 sum rule in Eq.~(\ref{eq:sumrule}), using  
the standard  PDG parameterization \cite{PDG} of the PMNS mixing
matrix. After this derivation of the  sum rule, 
and detailed discussion of the conditions of its
validity, we then apply it to present oscillation
data in Sec.~\ref{sec:fit}. Sec.~\ref{sec:future} is devoted to the
simulation of future experiments including long-baseline reactor
experiments, various second generation superbeam setups, a
$\beta$-beam, and neutrino factories. We summarize in
Sec.~\ref{sec:conclusions}.

%%%%%%%%%%%%%%%%%%%%%%%%%%%%%%%%%%%%%%%%%%%%%%%%%%%%%%%%%%%%%%%%%%% 
\section{Derivation of the  Sum Rule}
%%%%%%%%%%%%%%%%%%%%%%%%%%%%%%%%%%%%%%%%%%%%%%%%%%%%%%%%%%%%%%%%%%% 
\label{sec:sumrule}

The mixing matrix in the lepton sector, the PMNS matrix
$U_{\mathrm{PMNS}}$, is defined as the matrix which appears in the
electroweak coupling to the $W$ bosons expressed in terms of lepton
mass eigenstates. With the mass matrices of charged leptons
$M_\mathrm{e}$ and neutrinos $m_{\nu}$ written as\footnote{Although we
have chosen to write a Majorana mass matrix, all relations in the
following are independent of the Dirac or Majorana nature of neutrino
masses.}
\begin{eqnarray}
{\cal L}=-  \bar{e}_L M_\mathrm{e} e_R  
- \tfrac{1}{2}\bar{\nu}_L m_{\nu} \nu_\mathrm{L}^c 
+ \text{H.c.}\; ,
\end{eqnarray}
and performing the transformation from flavour to mass basis by
 \begin{eqnarray}\label{eq:DiagMe}
V_{\mathrm{e}_\mathrm{L}} \, M_{\mathrm{e}} \,
V^\dagger_{\mathrm{e}_\mathrm{R}} =
\mbox{diag}(m_e,m_\mu,m_\tau)
 , \quad
V_{\nu_\mathrm{L}} \,m_\nu\,V^T_{\nu_\mathrm{L}} =
\mbox{diag}(m_1,m_2,m_3),
\end{eqnarray}
the PMNS matrix is given by
\begin{eqnarray}\label{Eq:PMNS_Definition}
U_{\mathrm{PMNS}} 
= V_{e_\mathrm{L}} V^\dagger_{\nu_\mathrm{L}} \,.
\end{eqnarray}
Here it is assumed implicitly that unphysical phases are removed by
field redefinitions, and $U_\mathrm{PMNS}$ contains one Dirac phase
and two Majorana phases. The latter are physical only in the case of
Majorana neutrinos, for Dirac neutrinos the two Majorana phases can be
absorbed as well. 

Many attractive classes of models lead to interesting predictions for
the neutrino mass matrix $m_\nu$, such as for instance tri-bimaximal
\cite{tribi} or bimaximal \cite{bimax} mixing where
$V^\dagger_{\nu_\mathrm{L}}$ takes the forms
\begin{eqnarray}
V^\dagger_{\nu_\mathrm{L},\mathrm{tri}} =
\left(\begin{array}{ccc}
\!\sqrt{2/3}&1/\sqrt{3} &0\!\\
\!-1/\sqrt{6}&1/\sqrt{3}&1/\sqrt{2}\!\\
\!1/\sqrt{6}&-1/\sqrt{3}&1/\sqrt{2}\!
\end{array}
\right)\;\mbox{or}\;\;
V^\dagger_{\nu_\mathrm{L},\mathrm{bi}} = 
\left(\begin{array}{ccc}
\!1/\sqrt{2}& 1/\sqrt{2}&0\!\\
\!-1/2&1/2&1/\sqrt{2}\!\\
\!1/2&-1/2&1/\sqrt{2}\!
\end{array}
\right),
\end{eqnarray}
respectively, although the  sum rule is not necessarily restricted
to either of these two forms.
As mentioned in the introduction such predictions are
not directly experimentally accessible because of the presence of the
charged lepton corrections. However, this challenge can be overcome
when we make the additional assumption that the charged lepton mixing
matrix has a CKM-like structure, in the sense that $V_{e_\mathrm{L}}$
is dominated by a 1-2 mixing, i.e.\ that its elements
$(V_{e_\mathrm{L}})_{13}$, $(V_{e_\mathrm{L}})_{23}$,
$(V_{e_\mathrm{L}})_{31}$ and $(V_{e_\mathrm{L}})_{32}$ are
very small compared to $(V_{e_\mathrm{L}})_{ij}$ ($i,j = 1,2$). 
In the following simplified derivation, we shall take these elements to be 
approximately zero, i.e.\
\begin{eqnarray}\label{Eq:AssumptionForUe}
V_{e_\mathrm{L}} \approx 
\left(\begin{array}{ccc}
\!(V_{e_\mathrm{L}})_{11}& (V_{e_\mathrm{L}})_{12}&0\!\\
\!(V_{e_\mathrm{L}})_{21}& (V_{e_\mathrm{L}})_{22}&0\!\\
\!0&0&1\!
\end{array}
\right) ,
\end{eqnarray}
and later on comment on the effect of them being non-zero (see footnote \ref{theta12e}). For a derivation including these elements, see \cite{Antusch:2005kw}.
This situation arises in many generic classes of flavour models in
the context of unified theories of fundamental interactions, where
quarks and leptons are joined in representations of the unified gauge
symmetries~\cite{sumrule,reviews}. 

Under this assumption, it follows directly from
Eq.~(\ref{Eq:PMNS_Definition}) that $(U_{\mathrm{PMNS}})_{31}$,
$(U_{\mathrm{PMNS}})_{32}$ and $(U_{\mathrm{PMNS}})_{33}$ are
independent of $V_{e_\mathrm{L}}$, and depend only on the
diagonalization matrix $V^\dagger_{\nu_\mathrm{L}}$ of the neutrino
mass matrix. This leads to the parameterization-independent sum rules
which we give in this form for the first time:
\begin{subequations}\label{Eq:InvSumrules}\begin{eqnarray}
\label{Eq:InvSumrule1} |(V^\dagger_{\nu_\mathrm{L}})_{31}| &\approx& 
  |(U_{\mathrm{PMNS}})_{31}|\;,\\
\label{Eq:InvSumrule2} |(V^\dagger_{\nu_\mathrm{L}})_{32}| &\approx& 
  |(U_{\mathrm{PMNS}})_{32}|\;,\\
\label{Eq:InvSumrule3} |(V^\dagger_{\nu_\mathrm{L}})_{33}| &\approx&
  |(U_{\mathrm{PMNS}})_{33}|\;.
\end{eqnarray}\end{subequations} 
These innocuous looking 
relations enable powerful tests of the structure of the neutrino mass
matrix in the presence of charged lepton
corrections.
Note that the left-hand sides of these relations
involve neutrino mixing matrix elements in a particular basis,
whereas the right-hand sides are basis invariant quantities.
This makes sense in the framework of a flavour theory which
has a preferred basis, the so-called ``theory basis''.\footnote{Also
note that models of neutrino masses have a basis-invariant
classification. For example, models of tri-bimaximal neutrino mixing via Constrained Sequential Dominance (CSD) \cite{King:2005bj}
fall in an invariant class of seesaw models, even in the presence
of charged lepton corrections, as discussed in \cite{King:2006hn}.}

Let us now study the sum rules in the standard PDG parameterization of the
PMNS matrix (see e.g.\ \cite{PDG}), 
\begin{eqnarray}\label{Eq:StandardParametrization}
 %\hspace{-0.5cm} 
 U_{\mathrm{PMNS}} = \left(
  \begin{array}{ccc}
  c_{12}c_{13} & 
  s_{12}c_{13} & s_{13}e^{-\I\delta}\\
  -c_{23}s_{12}-s_{13}s_{23}c_{12}e^{\I\delta} &
  c_{23}c_{12}-s_{13}s_{23}s_{12}e^{\I\delta}  &
  s_{23}c_{13}\\
  s_{23}s_{12}-s_{13}c_{23}c_{12}e^{\I\delta} &
  -s_{23}c_{12}-s_{13}c_{23}s_{12}e^{\I\delta} &
  c_{23}c_{13}
  \end{array}
  \right) \, P_\mathrm{Maj} \, ,
\end{eqnarray}
which is used in most analyses of neutrino oscillation experiments.
Here $\delta$ is the so-called Dirac CP violating phase which is in
principle measurable in neutrino oscillation experiments, and 
$P_\mathrm{Maj} = \mathrm{diag}(e^{\I \tfrac{\alpha_1}{2}}, 
e^{\I\tfrac{\alpha_2}{2}}, 0)$ contains the Majorana phases $\alpha_1,
\alpha_2$.  In the following we will use this standard
parameterization (including additional phases) also for 
$V^\dagger_{\nu_\mathrm{L}}$ and denote the
corresponding mixing angles by $\theta_{ij}^\nu$, while the mixing
angles $\theta_{ij}$ without superscript refer to the PMNS matrix.

In addition to the assumption that $V_{e_\mathrm{L}}$ is of the form
of Eq.~(\ref{Eq:AssumptionForUe}) we will now assume that the 1-3
mixing in the neutrino mass matrix is negligible,
\begin{equation}\label{eq:th13nu}
(V^\dagger_{\nu_\mathrm{L}})_{13} \approx 0 \,.
\end{equation}
Many textures for the neutrino mass matrix fulfill this relation
exactly, for example the cases of bimaximal and tri-bimaximal mixing,
although the assumption in Eq.~(\ref{eq:th13nu}) is more general.
Using the assumption (\ref{eq:th13nu}) in the sum rule of
Eq.~(\ref{Eq:InvSumrule1}) one obtains
\begin{equation}\label{eq:sumrule_abs}
s_{23}^\nu s_{12}^\nu \approx
\left| s_{23}s_{12} - s_{13}c_{23}c_{12}e^{\I\delta} \right|
\approx 
s_{23}s_{12} - s_{13}c_{23}c_{12} \cos(\delta) \,,
\end{equation}
where the last step holds to leading order in
$s_{13}$. Furthermore, Eq.~(\ref{Eq:InvSumrule3}) together with
Eq.~(\ref{eq:th13nu}) implies
\begin{equation}\label{eq:th23rule}
\theta_{23}^\nu \approx 
\theta_{23} + \mathcal{O}(\theta_{13}^2) \,. 
\end{equation}
Using this relation in Eq.~(\ref{eq:sumrule_abs}) leads to the sum
rule
\begin{equation}\label{eq:sumrule-gen}
\theta_{12} - \theta_{13} \cot(\theta_{23}) \cos (\delta) 
\approx \theta_{12}^\nu\;,
\end{equation}
which holds up to first order in $\theta_{13}$. Hence, we have
obtained an approximate expression for the (in general unobservable)
mixing angle $\theta_{12}^\nu$ in terms of directly measurable
parameters of the PMNS matrix. This sum rule can be used to test a
bimaximal ($\theta^{\nu}_{12} = \tfrac{\pi}{4}$) or tri-bimaximal
($\theta^{\nu}_{12} = \arcsin (\tfrac{1}{\sqrt{3}})$) structure of the
neutrino mass matrix, but may as well be applied for a different
pattern beyond these two examples.\footnote{
We would like to remark at this point that the sum rule holds 
at low energy, where the neutrino oscillation experiments are performed. 
Therefore, if theory predictions arise at high energies like the
GUT scale, their renormalization group evolution has to be taken 
into account. In seesaw models, the running can be calculated 
conveniently using the software package REAP \cite{REAP}.}
In the following we will specialise our discussion to models
predicting maximal 2-3 mixing in the neutrino mass matrix,
$\theta_{23}^\nu = \tfrac{\pi}{4}$. This includes of course the cases
of bimaximal and tri-bimaximal mixing. With Eq.~(\ref{eq:th23rule})
this leads to the  sum rule of Eq.~(\ref{eq:sumrule})~\cite{sumrule}.

As a side remark we mention that under the above conditions also a
simple relation for $|(U_{\mathrm{PMNS}})_{13}|$ can be obtained, 
$|(U_{\mathrm{PMNS}})_{13}| \approx |(V_{e_\mathrm{L}})_{12}|\,
|(V^\dagger_{\nu_\mathrm{L}})_{23}|$. In the standard
parameterization it yields
\begin{eqnarray}\label{eq:th13rule}
\theta_{13} 
\approx s_{23} \, \theta^e_{12}  
\approx \tfrac{1}{\sqrt{2}} \theta^e_{12}  \;,
\end{eqnarray}
where $|(V_{e_\mathrm{L}})_{12}| = \sin(\theta^e_{12}) \approx
\theta^e_{12}$, and in the last step we have approximated
$\theta_{23}\approx \tfrac{\pi}{4}$.  Hence, $\theta_{13}$ is related
to the 1-2 mixing in the charged lepton mass matrix.\footnote{Note
that in the derivation of the sum rule Eq.~(\ref{eq:sumrule-gen}) it
was not necessary to assume that $\theta^e_{12}$ is small. This
requirement follows only {\it a posteriori} from
Eq.~(\ref{eq:th13rule}) and the fact that $\theta_{13}$ has to be
small from data.} This relation has been noticed by many authors,
e.g.\ Refs.~\cite{sumrule,refs1,refs2,refs7,refs3,refs5,refs6,refs4,reviews}, 
and it can provide
additional hints on the underlying theory of flavour (see for example
Ref.~\cite{sumrule}). Here we will not explore this relation further
but focus on the  sum rule (\ref{eq:sumrule}). Note, however, that
under the assumptions (\ref{Eq:AssumptionForUe}) and (\ref{eq:th13nu})
the only mixing parameters of the model are $\theta^e_{12},
\theta_{12}^\nu, \theta_{23}^\nu$, and through the relations
(\ref{eq:th13rule}), (\ref{eq:sumrule-gen}), (\ref{eq:th23rule}) all
of them can be expressed in terms of measurable PMNS parameters.

Finally, we mention that under the above assumptions
Eq.~(\ref{eq:th23rule}) can be used to test predictions for
$\theta_{23}^\nu$. This is complementary to the application of the 
sum rule for $\theta_{12}^\nu$, and a precise determination of
$\theta_{23}$ will allow for an additional test of predictions for the
neutrino mass matrix.\footnote{\label{theta12e} If the assumption of Eq.~(\ref{Eq:AssumptionForUe}) is
relaxed and one allows for a small (but non-zero) 2-3 mixing in the
charged lepton mixing matrix, $\theta_{23}^e \ll 1$, there will be a
correction of order $\theta_{23}^e$ to Eq.~(\ref{eq:th23rule}), which has
to be taken into account when drawing conclusions on $\theta_{23}^\nu$
from a measurement of $\theta_{23}$. It can be shown~\cite{sumrule},
however, that this correction does not affect the $\theta_{12}$ sum rule
(\ref{eq:sumrule-gen}) to leading order.
} 
For the examples of tri-bimaximal and bimaximal
neutrino mixing, one can test experimentally the prediction
$\theta_{23} \approx \tfrac{\pi}{4}$, in addition to the verification
of the corresponding  sum rule for $\theta_{12}$. Prospects for the
measurement of deviations from maximal 2-3 mixing have been discussed
e.g.\ in Ref.~\cite{Antusch:2004yx}.

%%%%%%%%%%%%%%%%%%%%%%%%%%%%%%%%%%%%%%%%%%%%%%%%%%%%%%%%%%%%%%%%%%%
\section{The  Sum Rule and Present Oscillation Data}
%%%%%%%%%%%%%%%%%%%%%%%%%%%%%%%%%%%%%%%%%%%%%%%%%%%%%%%%%%%%%%%%%%%
\label{sec:fit}

In this section we show that already with present global data from
neutrino oscillation experiments the  sum rule can be used to test the
hypotheses of bimaximal or tri-bimaximal mixing in the neutrino mass
matrix. Using the results from the global analysis of
Ref.~\cite{Maltoni:2004ei} a fit is performed under the assumption of
the  sum rule Eq.~(\ref{eq:sumrule}) with the constraints
$\theta^\nu_{12} = 45^\circ$ for bimaximal mixing or
$\theta^{\nu}_{12} = \arcsin (\tfrac{1}{\sqrt{3}}) \approx 35.3^\circ$  
for tri-bimaximal mixing. (See also Ref.~\cite{Masina:2005hf} for similar 
considerations.) 

\begin{figure}[t]
\begin{center}
 \ensuremath{\vcenter{\hbox{\includegraphics[scale=0.5]{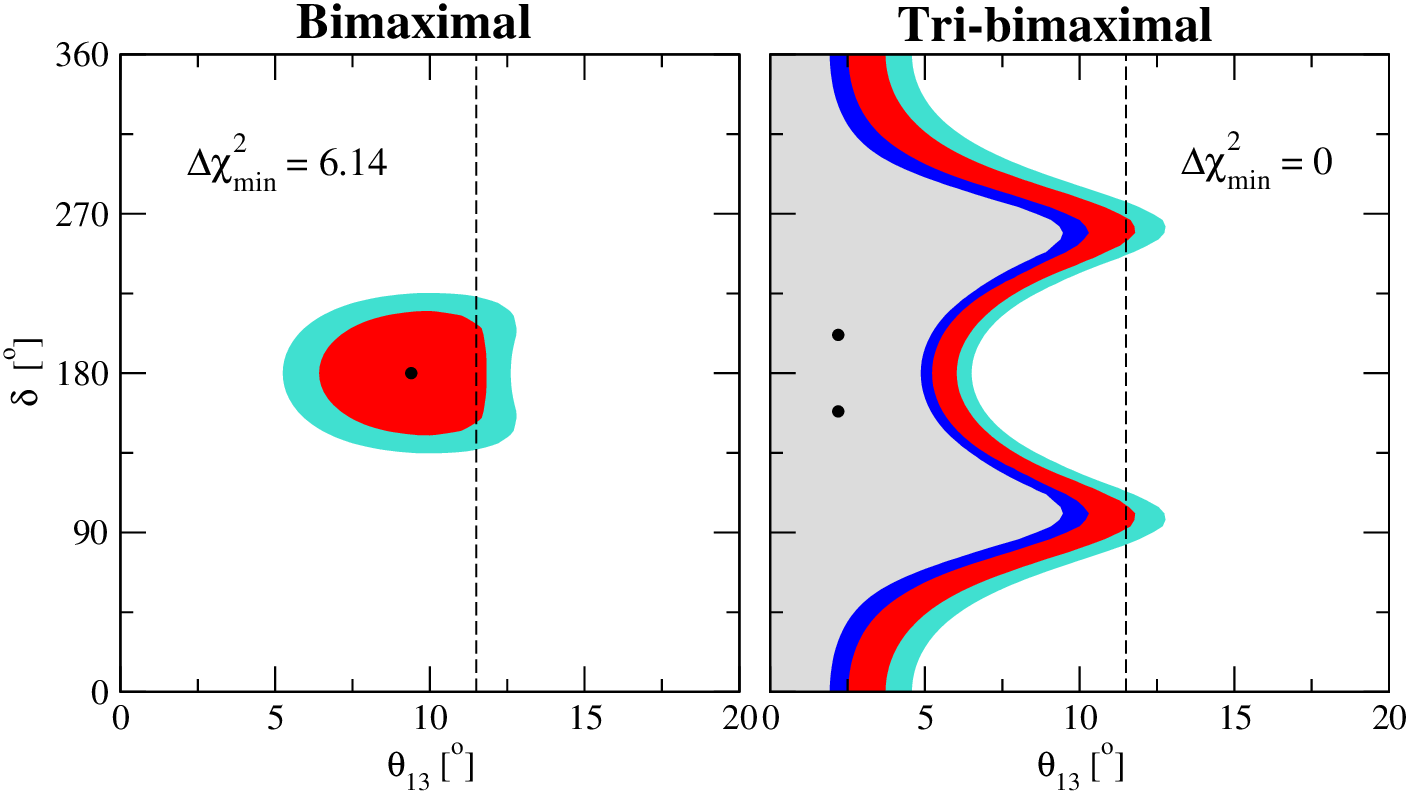}}}}
\vspace{0.2cm} 
  \mycaption{Fit to the present global data on neutrino oscillations
  under the assumption of the  sum rule and bimaximal (left) and
  tri-bimaximal (right) mixing in the neutrino mass matrix. Allowed
  regions are shown at 90\%, 95\%, 99\%, 99.73\%~CL (2 dof) with
  respect to the unconstrained best fit point. The dashed lines
  correspond to the upper bound on $\theta_{13}$ at $3\,\sigma$.}
  \label{fig:fit}
\end{center}
\end{figure}

Present data implies that $\theta_{12}$ is significantly smaller than
$45^\circ$, with the upper limit of $39.2^\circ$ at $3\,\sigma$
dominated by the SNO solar neutrino
experiment~\cite{Ahmed:2003kj}. Hence, to reconcile the value
$\theta^\nu_{12} = 45^\circ$ for bimaximal mixing one needs a
relatively large value of $\theta_{13}$ and $\cos(\delta) \simeq
-1$. These expectations are confirmed by the fit, as visible in the
left panel of Fig.~\ref{fig:fit}. We obtain $\Delta\chi^2_\mathrm{min}
\approx 6.14$ with respect to the unconstrained best fit point, and
hence, allowed regions appear only at 99\% and 99.73\%~CL. Fitting the
 sum rule for bimaximal mixing requires that both, $\theta_{12}$ and
$\theta_{13}$, are pushed towards their upper limits which leads to
the increase of $\chi^2$ mentioned above. We conclude that already
present data disfavours bimaximal neutrino mixing under the assumption of the
 sum rule at more than $2\sigma$. Hence, either the hypothesis of
$\theta_{12}^\nu = 45^\circ$ has to be discarded, or some of the
approximations needed for the  sum rule are not justified, for instance
the charged lepton corrections cannot be of CKM type as assumed in
Eq.~(\ref{Eq:AssumptionForUe}).  On the other hand, if the fit is
accepted, the  sum rule for bimaximal mixing predicts
that $\theta_{13}$ is close to its present bound and $\delta \simeq
\pi$.

The right panel of Fig.~\ref{fig:fit} shows the result for
the  sum rule with
tri-bimaximal neutrino mixing. In this case the fit is fully consistent with
the data and a best fit point at the same $\chi^2$ as the
unconstrained fit is found.
This follows since the best fit point $\theta_{12} =
33.2^\circ$ is close to the tri-bimaximal mixing value.
Indeed, for small values of $\theta_{13}$, say less than
$2^\circ$, the
 sum rule is satisfied within current experimental errors, for  
all values $\cos(\delta)$. On the other hand, the  sum rule
leads to a strengthening of the bound on $\theta_{13}$
in the regions where $\cos(\delta) \neq 0$.

%%%%%%%%%%%%%%%%%%%%%%%%%%%%%%%%%%%%%%%%%%%%%%%%%%%%%%%%%%%%%%%%%%%
\section{The  Sum Rule and Sensitivities of Future Experiments}
%%%%%%%%%%%%%%%%%%%%%%%%%%%%%%%%%%%%%%%%%%%%%%%%%%%%%%%%%%%%%%%%%%%
\label{sec:future}
 
In this section we explore the ability of future experiments to
constrain the parameter combination
$\theta_{12} - \theta_{13}\cos (\delta)$, 
appearing on the left-hand
side of the  sum rule in Eq.~(\ref{eq:sumrule}),
in order to obtain
information on $\theta_{12}^\nu$, and so enable a comparison with the predicted
values of $\theta_{12}^\nu$ coming from particular flavour models.
Obviously, in order to do this, the errors on
$\theta_{12}$ as well as on $\theta_{13}\cos(\delta)$ should be as
small as possible.

Let us first discuss the prospects to improve the accuracy on
$\theta_{12}$, which is $4.9^\circ$ at $3\,\sigma$ from present
data~\cite{Maltoni:2004ei}, dominated by the SNO solar neutrino
experiment~\cite{Ahmed:2003kj}. Significant improvement on
$\theta_{12}$ can be obtained by long-baseline (LBL) reactor neutrino
experiments, similar to the KamLAND experiment~\cite{Araki:2004mb}. 
A realistic possibility is that the Super-K experiment will be doped
with Gadolinium (SK-Gd)~\cite{Beacom:2003nk}. This will allow for a
very efficient detection of reactor $\bar\nu_e$, and by observing the
neutrino flux from the surrounding reactors in Japan a precise
determination of the ``solar'' oscillation parameters will be
obtained~\cite{Choubey:2004bf}. Following the analysis of
Ref.~\cite{Petcov:2006gy}, after 5 years of data taking an accuracy of
$4.0^\circ$ can be obtained for $\theta_{12}$ at $3\,\sigma$.
Another interesting option would be a big scintillator detector such as
LENA. In Ref.~\cite{Petcov:2006gy} the possibility of a 44~kt detector
installed in the Frejus underground laboratory has been considered. By
the observation of the reactor neutrino flux from nearby reactors in
France an accuracy of $2.0^\circ$ can be obtained for $\theta_{12}$ at
$3\,\sigma$ after 5 years of data taking.

Probably the best way to measure $\theta_{12}$ would be a dedicated
reactor experiment with only one reactor site at a distance around
$60\,\mathrm{km}$~\cite{Minakata:2004jt,Bandyopadhyay:2004cp}, where
the first survival probability minimum would be right in the middle of
the reactor event rate spectrum. This has been named ``SPMIN
experiment'' in Ref.~\cite{Bandyopadhyay:2004cp}. The obtainable
accuracy in this type of experiment, as in all reactor experiments, is
a balance between statistical and systematical errors. The former call
for large detectors and powerful reactors, whereas the latter require
great experimental skill and a careful design. For illustration we
consider here a rather ``big'' setup corresponding to an exposure of a
liquid scintillator detector of 200~GW~kt~y.\footnote{For comparison,
typical nuclear power plants have a thermal power output of order
10~GW, and the KamLAND experiment has a total mass of about 1~kt.}
The estimated accuracy at $3\,\sigma$ of such an experiment to
$\theta_{12}$ is $0.7^\circ$ from statistical errors only, and
$1.1^\circ$ if various systematical effects are taken into account.
These numbers have been obtained by applying a similar analysis for
the SPMIN experiment as in Ref.~\cite{Petcov:2006gy}, where also a
detailed description of the various systematics can be found.  At such
large exposures systematics have a big impact on the accuracy, but it
seems difficult to improve the systematics in a very large kiloton
sized detector.

\begin{table}[t]
  \centering
  \begin{tabular}{lc}
    \hline\hline
    & accuracy on $\theta_{12}$ at $3\,\sigma$ \\
    \hline
    present data                   & $4.9^\circ$ \\
    SK-Gd, 5 years                 & $4.0^\circ$ \\
    LENA @ Frejus, 44 kt, 5 y      & $2.0^\circ$ \\
    SPMIN @ 60 km, 200 GW kt y     & $1.1^\circ$ \\
    \hline\hline
  \end{tabular}
  \mycaption{Accuracy on $\theta_{12}$ from present data and three
  options for future LBL reactor experiments.}
  \label{tab:th12}
\end{table}

The reason why neither SK-Gd nor LENA at Frejus can compete with a
dedicated SPMIN experiment is that many nuclear reactors at various
distances contribute which washes out the oscillation signature to
some extent. Let us note that also future solar experiments, even with a
1\% measurement of the pp-flux, cannot compete with an SPMIN reactor
experiment~\cite{Bandyopadhyay:2004cp}.
The prospects of the $\theta_{12}$ measurement are summarized in
Tab.~\ref{tab:th12}. In the following combined analysis with LBL
accelerator experiments we will consider as reference values the
$3\,\sigma$ accuracies of $4.0^\circ, 2.0^\circ$ and $1.1^\circ$ obtainable
at SK-Gd, LENA at Frejus, and a
$200\,\mathrm{GW}\,\mathrm{kt}\,\mathrm{y}$ SPMIN experiment, respectively.

Next we turn to the sensitivity of LBL accelerator experiments to
the combination of physical parameters appearing on the left-hand
side of the  sum rule in Eq.~(\ref{eq:sumrule}). 
These experiments are sensitive to
$\delta$ and $\theta_{13}$ but have nearly no sensitivity to
$\theta_{12}$.  Therefore, we will use the input on $\theta_{12}$ from
LBL reactor experiments as described above and perform a combined
reactor plus accelerator analysis. We follow the general analysis
procedure as described in Ref.~\cite{Huber:2002mx} with the difference
that we now project onto the direction $\theta_{12} -
\theta_{13}\cos(\delta)$ in the parameter space. Thus, the obtained
results do include the errors and the correlations on $\theta_{13}$
and $\delta$ as well as the errors on $\theta_{12}$, $\Delta
m^2_{21}$, $\theta_{23}$, $\Delta m^2_{31}$ and the matter
density. Especially the correlation between $\theta_{13}$ and $\delta$
is crucial, since the relevant oscillation probability contains terms
which go as
\begin{equation}
\theta_{13} \sin(\delta)\quad\mathrm{and}\quad\theta_{13} \cos(\delta)\,.
\end{equation}
However, the $L/E$-dependence of these two terms is different and
hence experiments covering different $L/E$-ranges may have very
different sensitivities to $\theta_{13}\cos(\delta)$. For these
reasons the accuracy on the combination $\theta_{13}\cos(\delta)$
may be very different from the accuracy individually obtained on
$\theta_{13}$ and $\delta$. Therefore, a proper treatment and
inclusion of the correlation between $\theta_{13}$ and $\delta$ is
mandatory to obtain meaningful results. 

For the experiments discussed in the following the sensitivity to
either $\theta_{13} \sin(\delta)$ or $\theta_{13} \cos(\delta)$ is
dominated by the data from the appearance channels. There are two main
reasons for this: The $\sin(\delta)$ term can only appear in
off-diagonal transitions, {\it i.e.} appearance channels, because it
is manifestly CP violating.  Secondly, a possible $\cos(\delta)$
contribution in the disappearance channels is always suppressed with
respect to the leading $\theta_{23}$ effect and hence plays no
statistically significant role. Only for the very largest values of
$\theta_{13}$ there is contribution of the disappearance channels, but
it is still very small. We will not discuss the possible impact of
short baseline reactor experiments which are designed to determine
$\theta_{13}$. The reason is, that all the experiments discussed in
the following have a superior sensitivity to $\theta_{13}$ on their
own.

\begin{table}[t]
  \centering\small
  \begin{tabular}{lccll}
    \hline\hline
    Setup & Ref. & Baseline & Detector & Beam \\
    \hline
    SPL& \cite{Campagne:2006yx} & 130 km & 440 kt WC & 
          4 MW superbeam, 2 y ($\nu$) + 8 y ($\bar\nu$) \\
    T2HK& \cite{Campagne:2006yx} & 295 km & 440 kt WC & 
          4 MW superbeam, 2 y ($\nu$) + 8 y ($\bar\nu$) \\
    WBB& \cite{Barger:2006vy}  & 1300 km & 300 kt WC & 1.5 MW
    superbeam, 5 y ($\nu$) + 5 y ($\bar\nu$) \\
    BB350& \cite{Burguet-Castell:2005pa} & 730 km & 440 kt WC &
    $5\times1.1\cdot 10^{18}~^{18}$Ne  +    $5\times 2.9\cdot
    10^{18}~^{6}$He  \\
    NFC &\cite{Huber:2006wb} & 4000 km & 50 kt MID &50 GeV, $4\times10^{21}~\mu^-$ +  $4\times10^{21}~\mu^+$\\
    NFO &\cite{Huber:2006wb} & 4000+7500 km & 2$\times$50 kt MID* &20 GeV, $4\times10^{21}~\mu^-$ +  $4\times10^{21}~\mu^+$ \\
    \hline\hline
  \end{tabular}
  \mycaption{Summary of the six future LBL accelerator experiments
  considered in this study. WC stands for water \v{C}erenkov detector
  and all masses for this technology are fiducial masses. MID denotes
  a magnetized iron calorimeter, whereas MID* denotes an improved
  version thereof. In the column ``Beam'' we give for BB350 the total
  number of useful ion decays, and for NFC, NFO the energy of the
  stored muons and the total number of useful muon decays. For more
  details see the text.}
  \label{tab:setups}
\end{table}

The calculations are performed with the GLoBES software
package~\cite{Huber:2004ka}. For the input values of the oscillation
parameters we use $\Delta m^2_{31} = 2.5\times 10^{-3}$~eV$^2$,
$\Delta m^2_{21} = 8\times 10^{-8}$~eV$^2$, $\theta_{23} = 45^\circ$,
and $\theta_{12} = 33.2^\circ$. We consider six examples for future
experiments. Their main characteristics are summarized in
Tab.~\ref{tab:setups}. They include three second generation superbeam
experiments, SPL -- a CERN based experiment with a Mt size water
\v{C}erenkov detector at Frejus~\cite{Campagne:2006yx}, T2HK, the
second stage of the Japanese T2K project~\cite{Campagne:2006yx} (see
also Ref.~\cite{Huber:2002mx}), and WBB -- a wide-band beam with a
very long baseline as discussed in the
US~\cite{Barger:2006vy,Diwan:2003bp}.  Furthermore, we consider an
advanced $\beta$-beam setup BB350 as described in
Ref.~\cite{Burguet-Castell:2005pa}, with a relativistic
$\gamma$-factor of the decaying $^{18}$Ne and $^{6}$He ions of
350. All these experiment are planed to employ a large water
\v{C}erenkov detector with a fiducial mass in the range $300 - 440$
kt. Note, that the setup labeled WBB assumes an operational time per
solar year of $1.7\cdot10^{7}$ s instead of the usual $10^7$ s. The
two neutrino factory setups considered here, NFC and NFO are taken
from~\cite{Huber:2006wb}. NFC is what we call conservative, in the
sense that it employs only one magnetized iron detector (MID) with the
canonical properties regarding muon detection threshold and background
rejection~\cite{Cervera:2000vy,Huber:2002mx}.  NFO is an optimized
version, which uses two identical detectors at two baselines of 4000
km and 7500 km, the latter being the so-called magic
baseline~\cite{Huber:2003ak}. The second difference is that the
detector is now an improved MID*, which has a lower muon detection
threshold but somewhat larger backgrounds, for details
see~\cite{Huber:2006wb}. The lower threshold allows to reduce the muon
energy from 50 GeV to 20 GeV.

\begin{figure}[t]
\begin{center}
\ensuremath{\vcenter{\hbox{\includegraphics[width=0.95\textwidth]{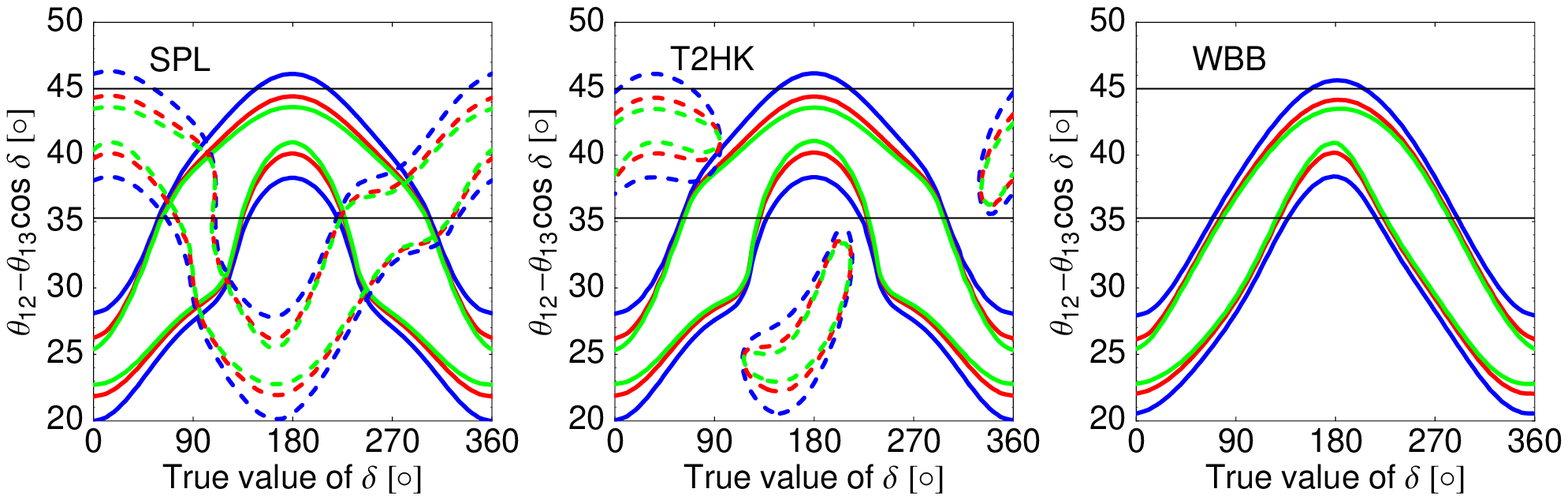}}}}
\ensuremath{\vcenter{\hbox{\includegraphics[width=0.95\textwidth]{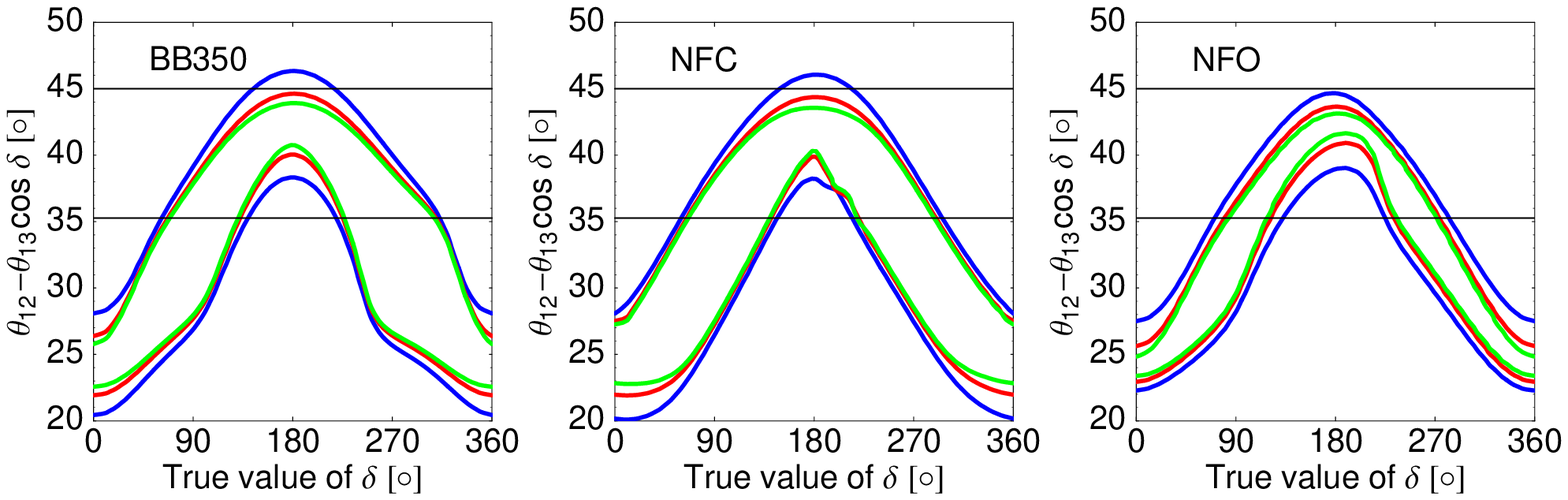}}}}
\vspace{0.2cm} 
  \mycaption{The $3\,\sigma$ allowed interval for the parameter
   combination $\theta_{12} - \theta_{13}\cos(\delta)$ 
appearing on the left-hand
side of the  sum rule in Eq.~(\ref{eq:sumrule}),
as a function of the true value of $\delta$ for
   $\sin^22\theta_{13}=10^{-1}$ from various LBL experiments. The
   dashed lines correspond to the $\mathrm{sgn}(\Delta m^2_{31})$
   degenerate solution. The colors indicate different errors on
   $\theta_{12}$: blue -- $4.0^\circ$, red -- $2.0^\circ$ and green --
   $1.1^\circ$ (at $3\,\sigma$). A true value
   $\sin^2 \theta_{12}=0.3$ ($\theta_{12}=33.2^\circ$) has been
   assumed. The horizontal lines show the  sum rule 
predictions corresponding to bimaximal and
   tri-bimaximal neutrino mixing.}
\label{fig:lth13}
\end{center}
\end{figure}

\begin{figure}[t]
\begin{center}
\ensuremath{\vcenter{\hbox{\includegraphics[width=0.95\textwidth]{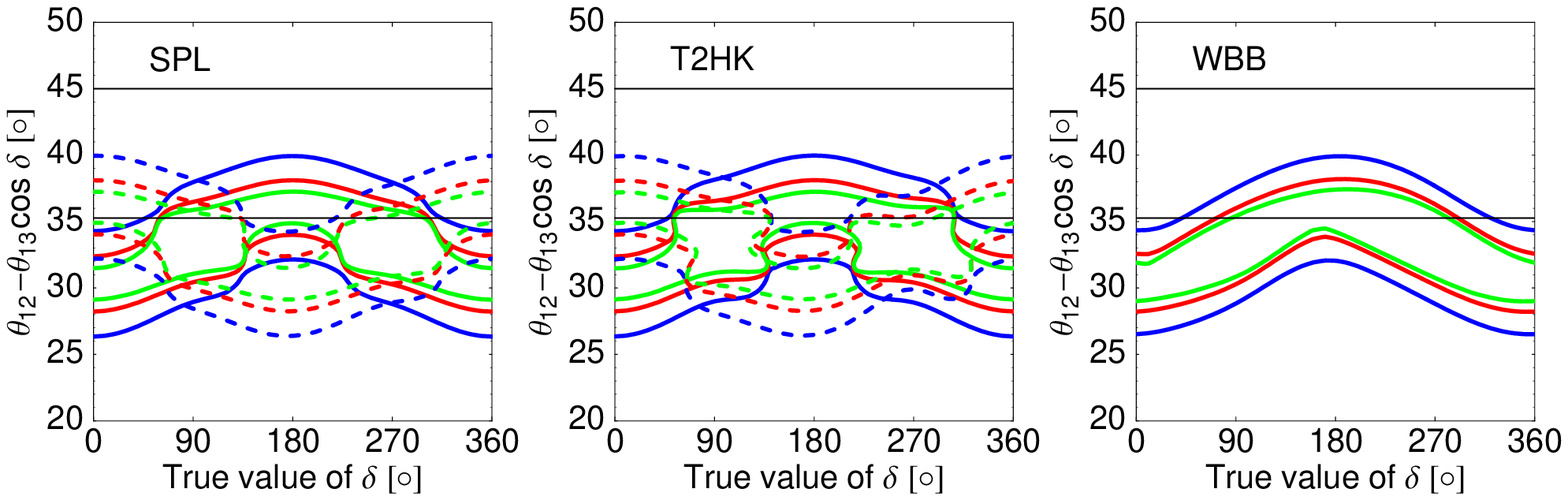}}}}
\ensuremath{\vcenter{\hbox{\includegraphics[width=0.95\textwidth]{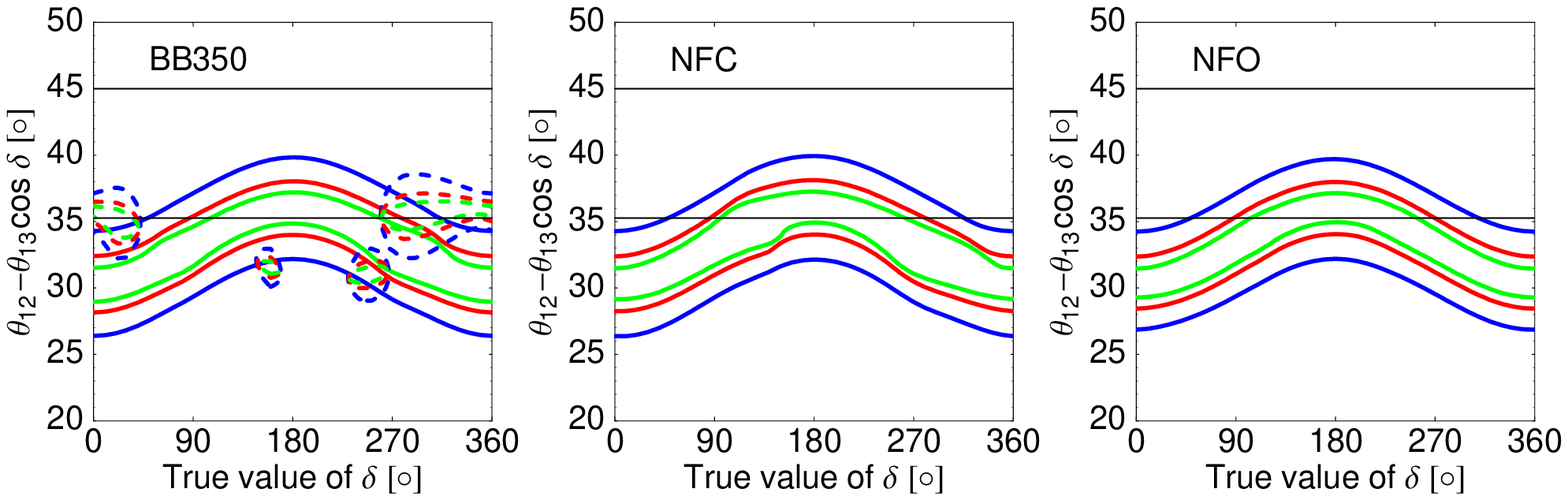}}}}
\vspace{0.2cm}
   \mycaption{Same as Fig.~\ref{fig:lth13} but for $\sin^22\theta_{13}=10^{-2}$.}
\label{fig:mth13}
\end{center}
\end{figure}

Figs.~\ref{fig:lth13} and \ref{fig:mth13} show the results for the
$3\,\sigma$ allowed interval in $\theta_{12} -
\theta_{13}\cos(\delta)$ as a function of the true value of $\delta$
assuming $\sin^22\theta_{13}=10^{-1}$ and $10^{-2}$, respectively,
from the considered experimental setups. This allowed interval can be
compared with theoretical predictions for $\theta_{12}^\nu$. We
illustrate in the figures the cases of bimaximal and tri-bimaximal
mixing by the horizontal lines, but of course any prediction for
$\theta_{12}^\nu$ can be confronted with the outcome of the
experiments.
Since we have used as true value for $\theta_{12}$ the present best
fit point of $33.2^\circ$, bimaximal mixing ($\theta_{12}^\nu =
45^\circ$) can be obtained only for large values of $\theta_{13}$ and
$\delta \simeq 180^\circ$, in agreement with the discussion in
Sec.~\ref{sec:fit}.  For larger (smaller) true values of
$\theta_{12}$, the bands and islands in Figs.~\ref{fig:lth13} and
\ref{fig:mth13} are shifted up (down) correspondingly.

All experiments shown in Figs.~\ref{fig:lth13} and \ref{fig:mth13}
have good sensitivity to $\theta_{12} - \theta_{13}\cos(\delta)$.  In
many cases only some specific values of the CP phase $\delta$ are
consistent with a given prediction for $\theta_{12}^\nu$, which
illustrates the power of the  sum rule. An interesting observation is
that the presence of the mass hierarchy degenerate solutions (dashed
lines) limits the usefulness of SPL and T2HK severely. In these
experiments the matter effect is small because of the relatively short
baseline. This implies that the mass hierarchy degenerate solution
cannot be resolved. Furthermore, the degenerate solution appears at a
similar value of $\theta_{13}$ but at a fake CP phase close to $\pi -
\delta$~\cite{Minakata:2001qm}. This changes the sign of the term
$\theta_{13}\cos(\delta)$, which explains the shape of the dashed
curves in the figures. Because of this degeneracy an ambiguity appears
when the  sum rule is applied for SPL and T2HK, which significantly
limits the possibility to distinguish between various predictions for
$\theta_{12}^\nu$, especially for large values of $\theta_{13}$ as
visible in Fig.~\ref{fig:lth13}. A solution to this problem could be
the information provided by atmospheric neutrinos in the Mt size
detectors used in these experiments~\cite{Huber:2005ep} (which is not
included here). For the other experiments the problem of the
degeneracy is absent, since the mass hierarchy degeneracy can be
resolved (at sufficiently large $\theta_{13}$) thanks to the longer
baselines.

\begin{figure}[t]
\begin{center}
\ensuremath{\vcenter{\hbox{\includegraphics[width=0.95\textwidth]{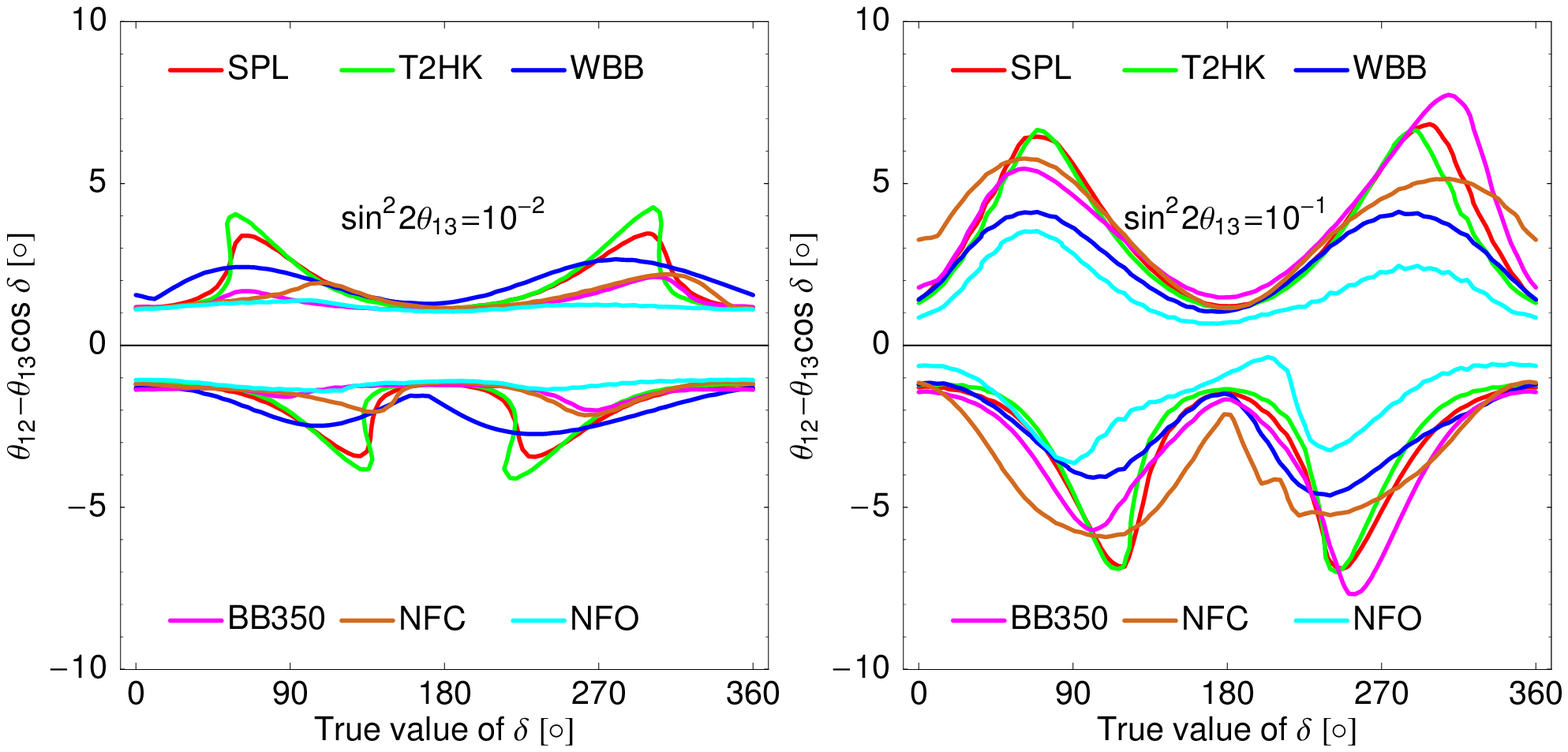}}}}
\vspace{0.5cm} 
  \mycaption{The $3\,\sigma$ error in degrees for the combination of
  parameters $\theta_{12} - \theta_{13}\cos(\delta)$, appearing on the
  left-hand side of the sum rule in Eq.~(\ref{eq:sumrule}), as a
  function of the true value of $\delta$ for
  $\sin^22\theta_{13}=10^{-2}$ (left) and $\sin^22\theta_{13}=10^{-1}$
  (right). The different colored lines are for different experiments
  as given in the legend. The $\mathrm{sgn}(\Delta m^2_{31})$
  degenerate solution has been omitted. Although $\sin^2 \theta_{12}=0.3$
($\theta_{12}=33.2^\circ$) has been used as true value, the results are practically
independent of this assumption. The error on $\theta_{12}$ is
  $1.1^\circ$ at $3\,\sigma$.}
\label{fig:error}
\end{center}
\end{figure}

The performance of all experiments is summarized also in
Fig.~\ref{fig:error}, which shows the obtainable $3\,\sigma$ accuracy
for the combination of parameters
$\theta_{12} - \theta_{13}\cos(\delta)$, 
appearing on the left-hand
side of the  sum rule in Eq.~(\ref{eq:sumrule}), as
a function of the true value of $\delta$ for the two cases
$\sin^22\theta_{13}=10^{-2}$ and $10^{-1}$.  
This figure shows that it
will be possible to discriminate between models whose predictions for
$\theta_{12}^\nu$ differ by a few degrees.
For the large value of $\theta_{13}$ assumed in Figs.~\ref{fig:lth13}
and \ref{fig:error} (right), $\sin^22\theta_{13}=10^{-1}$, the total
uncertainty is dominated by the term $\theta_{13}\cos(\delta)$ in the
 sum rule, and a modest improvement of the current error on
$\theta_{12}$ will be enough for exploring the  sum rule. The accuracy
depends significantly on the true value of $\delta$. Obviously the
impact of the term $\theta_{13}\cos(\delta)$ is larger for
$|\cos(\delta)| \simeq 1$. For smaller values of $\theta_{13}$ the
accuracy on $\theta_{12}$ becomes more important, the overall
sensitivity is dominated by the LBL reactor measurement, and the
dependence on $\delta$ is weaker.\footnote{As visible in 
Fig.~\ref{fig:mth13}, for a few
values of $\delta$ for T2HK the allowed region of $\theta_{12} -
\theta_{13}\cos(\delta)$ consists of two disconnected intervals even for
fixed neutrino mass hierarchy, because of the so-called intrinsic
degeneracy. This explains the ``turn over'' of the T2HK lines at some
values of $\delta$ in Fig.~\ref{fig:error} (left).}

It follows from Figs.~\ref{fig:lth13}, \ref{fig:mth13}, and
\ref{fig:error} that NFO has the best performance for all values of
$\theta_{13}$, making this the machine of choice for testing the  sum
rule.
NFC compares well to BB350 for this measurement, whereas the
performance on $\delta$ and $\theta_{13}$ individually is much worse
for NFC than for BB350. The reason for this
behaviour is that an experiment whose events are centered around the
first oscillation maximum like a $\beta$-beam or superbeam is mainly
sensitive to the $\theta_{13}\sin(\delta)$ term. A neutrino factory,
however, gets most of its events above the first oscillation maximum
and thus is much more sensitive to the $\theta_{13}\cos(\delta)$ term.
This explains also the relatively good performance of the WBB visible
from the right panel of Fig.~\ref{fig:error}, where WBB performs
second only to NFO. For such large values of $\theta_{13}$
($\sin^22\theta_{13}=10^{-1}$) spectral information far beyond the
first oscillation maximum can be explored efficiently, which is
important for constraining $\theta_{13}\cos(\delta)$.

For the somewhat smaller value of $\theta_{13}$,
$\sin^22\theta_{13}=10^{-2}$, the performances of NFO, NFC, and BB350
become rather similar, whereas the accuracies obtainable at superbeams
depend still to some extent on the true value of $\delta$, see
Fig.~\ref{fig:error} (left). Note that in this figure the most
optimistic accuracy on $\theta_{12}$ from an SPMIN reactor experiment
has been assumed, and that the mass hierarchy degeneracy has not been
taken into account. Indeed, decreasing the true value of $\theta_{13}$
for all setups besides the neutrino factory one, at some point the mass
hierarchy degenerate solution kicks in and introduces an ambiguity in
the allowed interval for $\theta_{12} - \theta_{13}\cos(\delta)$,
compare also Fig.~\ref{fig:mth13}.

\begin{figure}[t]
\begin{center}
  \includegraphics[width=0.5\textwidth]{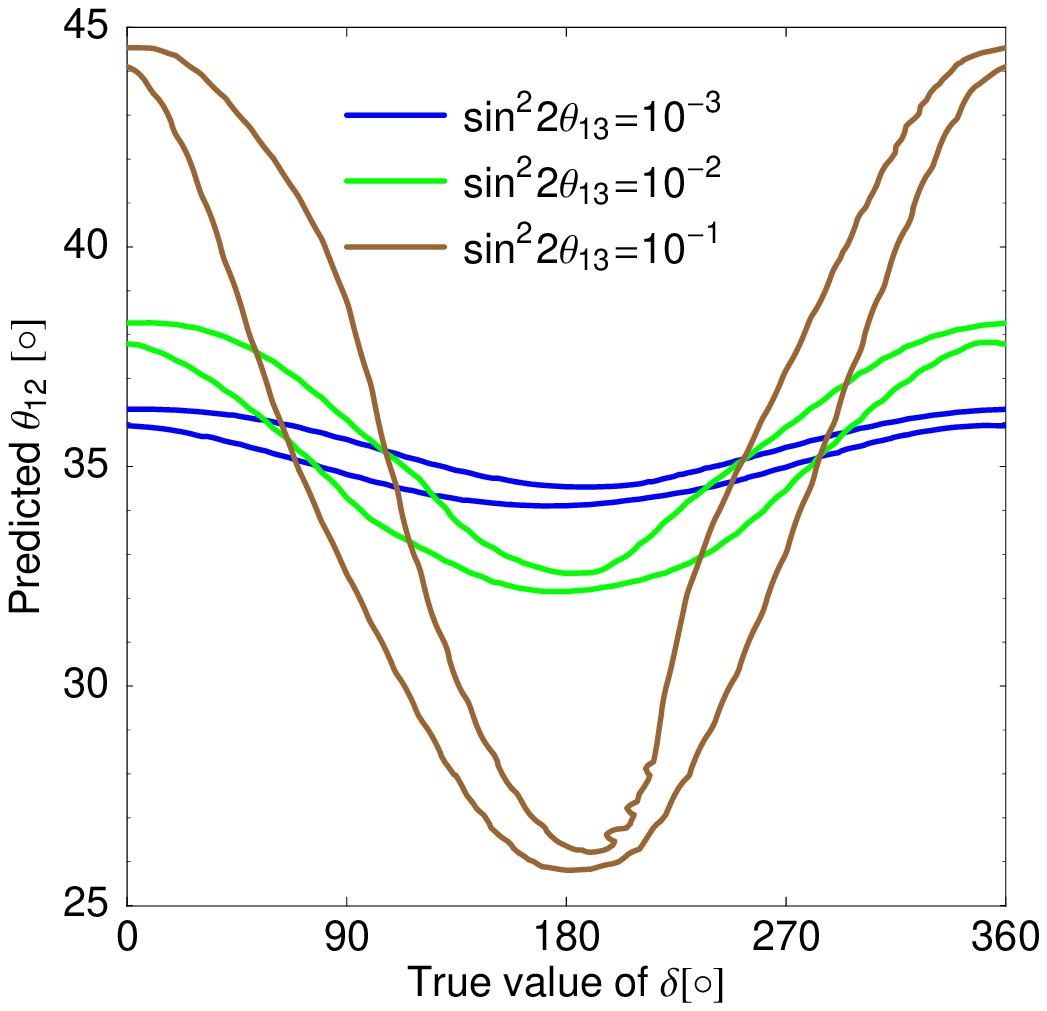}
  \mycaption{The prediction for $\theta_{12}$ from the  sum rule
  with tri-bimaximal neutrino mixing, as given in
  Eq.(\ref{eq:th12pred}). The different curves show the predicted
  $3\,\sigma$ interval for $\theta_{12}$ following from a NFO
  measurement of $\theta_{13}\cos(\delta)$ as a function of the true
  values of $\delta$ and $\theta_{13}$.} \label{fig:th12pred}
\end{center}
\end{figure}

For a given model prediction of the neutrino mixing angle 
$\theta^{\nu}_{12}$, the  sum rule 
in Eq.~(\ref{eq:sumrule}) may be expressed as 
a prediction for the physical solar mixing angle as a function of 
the CP violating Dirac oscillation phase $\delta$.
Fig.~\ref{fig:th12pred} shows the  sum rule prediction for the PMNS
parameter $\theta_{12}$ corresponding to 
tri-bimaximal mixing in the neutrino mass matrix,
$\theta^{\nu}_{12} = \arcsin (\tfrac{1}{\sqrt{3}}) \approx
35.3^\circ$, i.e.,
\begin{equation}\label{eq:th12pred}
\theta_{12} \approx 35.3^\circ + \theta_{13}\cos(\delta) \,.
\end{equation}
In the figure we have simulated data for the NFO setup for
different true values of $\theta_{13}$ and $\delta$ and used
Eq.~(\ref{eq:th12pred}) to calculate the resulting $3\,\sigma$ range
for the predicted $\theta_{12}$. This result can be compared with the
outcome of a separate measurement of $\theta_{12}$ (for example in a
reactor experiment) to test whether the hypothesis of tri-bimaximal
neutrino mixing is compatible with the assumptions leading to the sum
rule.

%%%%%%%%%%%%%%%%%%%%%%%%%%%%%%%%%%%%%%%%%%%%%%%%%%%%%%%%%
\section{Summary and Conclusions}
%%%%%%%%%%%%%%%%%%%%%%%%%%%%%%%%%%%%%%%%%%%%%%%%%%%%%%%%%
\label{sec:conclusions}

In this work we have considered the  sum rule in Eq.~(\ref{eq:sumrule}), 
and in particular how well the combination of parameters
$\theta_{12}- \theta_{13}\cos(\delta)$, which appears on the 
left-hand side, can be measured in oscillation experiments.
This is important, since the  sum rule follows from 
quite general assumptions which are satisfied in a wide
class of flavour models. Moreover, particular such flavour models
make definite predictions for $\theta_{12}^\nu$, and the  sum
rule then enables these models to be tested.

We have derived the  sum rule, starting from a parameterization
independent set of sum rules, which follow from certain well defined
assumptions about the nature of charged lepton and neutrino mixings.
We then expressed the  sum rule in terms of the standard PMNS mixing
parameters (see e.g.\ \cite{PDG}) commonly used in presenting the 
results of neutrino oscillation experiments. 
One way to view the  sum rule is to
consider the charged lepton corrections to the neutrino mixing angle
$\theta_{12}^\nu$ predicted from theory, leading to the physical solar neutrino
mixing angle $\theta_{12}$.  Then, under certain
assumptions, the charged lepton correction turns out to only depend on
the physical combination $\theta_{13}\cos(\delta)$.  To be precise,
the  sum rule in Eq.~(\ref{eq:sumrule}) holds up to first order in
$\theta_{13}$ under the following assumptions:
\begin{itemize}
\item[({\it a})]
The charged lepton mixing matrix is CKM-like, i.e., dominated by the
1-2 mixing angle, see Eq.~(\ref{Eq:AssumptionForUe}).
\item[({\it b})]
The 1-3 element of the neutrino mixing matrix is negligible, 
$\theta_{13}^\nu \approx 0$.
\item[({\it c})] The 2-3 mixing in the neutrino mass matrix is
maximal, $\theta_{23}^\nu \approx 45^\circ$, which under the previous
two assumptions is equivalent to $\theta_{23} \approx 45^\circ +
\mathcal{O}(\theta_{13}^2)$.
\end{itemize}
If condition ({\it c}) is not satisfied and $\theta_{23}$ turns out to
be non-maximal the generalized sum rule Eq.~(\ref{eq:sumrule-gen}) has
to be used. This would not change the reasoning of this paper, since
$\theta_{12}^\nu$ still can be expressed only in terms of measurable
quantities, involving now also $\theta_{23}$. Hence, condition
({\it c}) has been adopted here just for simplicity.
Condition ({\it a}) holds for a large class of models. For example, in
many GUT models the charged lepton mixing matrix is related to the
quark mixing matrix since quarks and leptons are joined in
representations of the unified gauge symmetries. On the other hand,
conditions ({\it b}) and ({\it c}) are rather typical for flavour
models in the neutrino sector. In particular, the popular examples of
bimaximal and tri-bimaximal mixing fulfill conditions ({\it b}) and
({\it c}) exactly.

We have demonstrated the usefulness of the  sum rule by imposing it as
a constraint in a fit to present global data from neutrino oscillation
experiments under the assumptions of bimaximal and tri-bimaximal
neutrino mixing. This analysis shows that under the condition ({\it
a}) bimaximal neutrino mixing is disfavoured at about $2\,\sigma$ by
present data with respect to tri-bimaximal mixing, which is perfectly
compatible with the data. If the fit for bimaximal mixing is accepted
the  sum rule predicts that $\theta_{13}$ is close to its present bound
and $\delta \simeq \pi$.

In the main part of the paper we have concentrated on 
the  sum rule in the context of future high
precision neutrino oscillation experiments. We have considered
long-baseline reactor experiments for a precise measurement of
$\theta_{12}$, as well as six examples for advanced long-baseline
accelerator experiments to constrain the parameter combination
$\theta_{13}\cos(\delta)$ appearing in the  sum rule. These setups
include three options for second generation superbeam experiments, a
$\beta$-beam, and two examples for a neutrino factory. It is shown
that most of these experiments will allow for a rather precise testing
of the  sum rule, and $\theta_{12}^\nu$ can be inferred within an
accuracy of few degrees, where the precise value shows some dependence
on the true values of $\theta_{13}$ and $\delta$. For
$\sin^22\theta_{13} \lesssim 10^{-2}$ the accuracy is dominated by the
error on $\theta_{12}$, whereas for large values of $\theta_{13}$ the
precision on the term $\theta_{13}\cos(\delta)$ dominates. Obviously
its impact is larger for $|\cos(\delta)| \simeq 1$. Because of the
appearance of $\cos(\delta)$ experiments operating not only at the
first oscillation maximum (where there is good sensitivity to
$\sin(\delta)$) are well suited for this kind of measurement, for
example a neutrino factory or a wide-band superbeam. 
Another interesting observation is that the mass hierarchy degeneracy
plays an important role. Since this degeneracy introduces an ambiguity
in the CP phase $\delta$ its appearance significantly reduces the
information on $\theta_{12}^\nu$ which can be extracted from the sum
rule.

To conclude, the  neutrino mixing sum rule considered in this work is a
convenient tool to explore the structure of the neutrino mass matrix
in the presence of charged lepton corrections, and to test whole
classes of models of neutrino masses and mixings. Already applied to
present data it is possible to obtain non-trivial statements, whereas
with future high precision oscillation experiments a rather accurate
testing of models will become possible in the framework of the  sum
rule.

\subsection*{Acknowledgments}

We would like to thank Michal Malinsky for reading the manuscript.  
The work of S.~Antusch was supported by the EU 6th Framework Program
MRTN-CT-2004-503369 ``The Quest for Unification: Theory Confronts
Experiment''. S.~F.~King and T.~Schwetz acknowledge the EU ILIAS project under contract 
RII3-CT-2004-506222 for support. 
Computations were performed on facilities supported by the NSF 
under Grants No.\ EIA-032078 (GLOW), PHY-0516857 (CMS Reserach 
Program subcontract from UCLA), and PHY-0533280 (DISUN), and by 
the WARF.

\providecommand{\bysame}{\leavevmode\hbox to3em{\hrulefill}\thinspace}


\begin{thebibliography}{10}

\bibitem{sumrule}
%\bibitem{King:2005bj}
  S.~F.~King,
  % ``Predicting neutrino parameters from SO(3) family symmetry and  quark-lepton
  %unification,''
  JHEP {\bf 0508} (2005) 105
  [arXiv:hep-ph/0506297]; 
%
%\bibitem{Masina:2005hf}
  I.~Masina,
  %``A maximal atmospheric mixing from a maximal CP violating phase,''
  Phys.\ Lett.\  B {\bf 633} (2006) 134
  [arXiv:hep-ph/0508031];
  %%CITATION = PHLTA,B633,134;%% 
%
%\bibitem{Antusch:2005kw}
  S.~Antusch and S.~F.~King,
%   ``Charged lepton corrections to neutrino mixing angles and CP phases
%  revisited,''
  Phys.\ Lett.\ B {\bf 631} (2005) 42
  [arXiv:hep-ph/0508044].  

\bibitem{PDG} 
W.-M.~Yao {\it et al.}\ [Particle Data Group Collaboration], 
J.\ Phys.\ G {\bf 33} (2006) 1.
    
\bibitem{tribi}
P.~F.~Harrison, D.~H.~Perkins and W.~G.~Scott,
%``Tri-bimaximal mixing and the neutrino oscillation data,''
Phys.\ Lett.\ B {\bf 530} (2002) 167
[arXiv:hep-ph/0202074].
%
A similar but physically different 
form was proposed earlier in: L.~Wolfenstein,
%``Oscillations Among Three Neutrino Types And CP Violation,''
Phys.\ Rev.\ D {\bf 18} (1978) 958.

\bibitem{bimax}
  V.~D.~Barger, S.~Pakvasa, T.~J.~Weiler and K.~Whisnant,
  %``Bi-maximal mixing of three neutrinos,''
  Phys.\ Lett.\ B {\bf 437} (1998) 107
  [arXiv:hep-ph/9806387].

\bibitem{refs1}
See e.g.: 
%\bibitem{Ohlsson:2002rb}
T.~Ohlsson and G.~Seidl,
  %``A flavor symmetry model for bilarge leptonic mixing and the lepton
  %masses,''
  Nucl.\ Phys.\ B {\bf 643} (2002) 247
  [arXiv:hep-ph/0206087];
  %%CITATION = HEP-PH 0206087;%%
C.~Giunti and M.~Tanimoto,
  %``Deviation of neutrino mixing from bi-maximal,''
  Phys.\ Rev.\ D {\bf 66}, 053013 (2002)
  [arXiv:hep-ph/0207096].

\bibitem{refs2}
S.~F.~King,
  %``Constructing the large mixing angle MNS matrix in see-saw models with
  %right-handed neutrino dominance,''
  JHEP {\bf 0209} (2002) 011
  [arXiv:hep-ph/0204360]; 
C.~Giunti and M.~Tanimoto,
  %``CP violation in bilarge lepton mixing,''
  Phys.\ Rev.\ D {\bf 66}, 113006 (2002)
  [arXiv:hep-ph/0209169];
  %%CITATION = HEP-PH 0209169;%%
P.~H.~Frampton, S.~T.~Petcov and W.~Rodejohann,
  %``On deviations from bimaximal neutrino mixing,''
  Nucl.\ Phys.\ B {\bf 687} (2004) 31
  [arXiv:hep-ph/0401206];
  %%CITATION = HEP-PH 0401206;%%
G.~Altarelli, F.~Feruglio and I.~Masina,
  %``Can neutrino mixings arise from the charged lepton sector?,''
  Nucl.\ Phys.\ B {\bf 689} (2004) 157
  [arXiv:hep-ph/0402155];
A.~Romanino,
  %``Charged lepton contributions to the solar neutrino mixing and
  %theta(13),''
  Phys.\ Rev.\ D {\bf 70}, 013003 (2004)
  [arXiv:hep-ph/0402258].
  %%CITATION = HEP-PH 0402258;%%

\bibitem{refs7} 
K.~A.~Hochmuth and W.~Rodejohann,
  %``Low and high energy phenomenology of quark-lepton complementarity
  %scenarios,''
  arXiv:hep-ph/0607103.
  %%CITATION = HEP-PH 0607103;%% 

\bibitem{refs3}
F.~Feruglio,
  %``Models of neutrino masses and mixings,''
  Nucl.\ Phys.\ Proc.\ Suppl.\  {\bf 143}, 184 (2005)
  [Nucl.\ Phys.\ Proc.\ Suppl.\  {\bf 145}, 225 (2005)]
  [arXiv:hep-ph/0410131];
  %%CITATION = HEP-PH 0410131;%%
Z.~Z.~Xing,
  %``Nontrivial correlation between the CKM and MNS matrices,''
  Phys.\ Lett.\ B {\bf 618} (2005) 141
  [arXiv:hep-ph/0503200].
  %%CITATION = HEP-PH 0503200;%%
   
\bibitem{refs5}   
F.~Plentinger and W.~Rodejohann,
  %``Deviations from tribimaximal neutrino mixing,''
  Phys.\ Lett.\ B {\bf 625} (2005) 264
  [arXiv:hep-ph/0507143].
  %%CITATION = HEP-PH 0507143;%%

\bibitem{refs6} 
R.~N.~Mohapatra and W.~Rodejohann,
  %``Broken mu-tau symmetry and leptonic CP violation,''
  Phys.\ Rev.\ D {\bf 72} (2005) 053001
  [arXiv:hep-ph/0507312].
  %%CITATION = HEP-PH 0507312;%%

\bibitem{refs4} 
T.~Ohlsson,
  %``Bimaximal fermion mixing from the quark and leptonic mixing matrices,''
  Phys.\ Lett.\ B {\bf 622} (2005) 159
  [arXiv:hep-ph/0506094].
  %%CITATION = HEP-PH 0506094;%%

\bibitem{Antusch:2005kw}
  S.~Antusch and S.~F.~King, in \cite{sumrule}.

\bibitem{reviews}
Further examples of models can, for example, be found in:
%\bibitem{King:2003rf}
  S.~F.~King and G.~G.~Ross,
  %``Fermion masses and mixing angles from SU(3) family symmetry and
  %unification,''
  Phys.\ Lett.\ B {\bf 574} (2003) 239
  [arXiv:hep-ph/0307190];
  %%CITATION = HEP-PH 0307190;%%
%\bibitem{deMedeirosVarzielas:2005ax}
  I.~de Medeiros Varzielas and G.~G.~Ross,
  %``SU(3) family symmetry and neutrino bi-tri-maximal mixing,''
  Nucl.\ Phys.\ B {\bf 733} (2006) 31
  [arXiv:hep-ph/0507176];
  %%CITATION = HEP-PH 0507176;%%
%\bibitem{deMedeirosVarzielas:2006fc}
  I.~de Medeiros Varzielas, S.~F.~King and G.~G.~Ross,
  %``Neutrino tri-bi-maximal mixing from a non-Abelian discrete family
  %symmetry,''
  arXiv:hep-ph/0607045;
  %%CITATION = HEP-PH 0607045;%%
S.~F.~King and M.~Malinsky,
  %``Towards a complete theory of fermion masses and mixings with SO(3) family
  %symmetry and 5d SO(10) unification,''
  JHEP {\bf 0611} (2006) 071
  [arXiv:hep-ph/0608021];
S.~F.~King and M.~Malinsky,
  %``A(4) family symmetry and quark-lepton unification,''
  arXiv:hep-ph/0610250. 

\bibitem{King:2005bj}
S.~F.~King, in \cite{sumrule}.

\bibitem{King:2006hn}
S.~F.~King,
%``Invariant see-saw models and sequential dominance,''
arXiv:hep-ph/0610239.
%%CITATION = HEP-PH 0610239;%%

\bibitem{REAP}
  S.~Antusch, J.~Kersten, M.~Lindner, M.~Ratz and M.~A.~Schmidt,
  %``Running neutrino mass parameters in see-saw scenarios,''
  JHEP {\bf 0503} (2005) 024
  [arXiv:hep-ph/0501272].
  %%CITATION = JHEPA,0503,024;%%

\bibitem{Antusch:2004yx}
  S.~Antusch, P.~Huber, J.~Kersten, T.~Schwetz and W.~Winter,
  %``Is there maximal mixing in the lepton sector?,''
  Phys.\ Rev.\ D {\bf 70} (2004) 097302
  [arXiv:hep-ph/0404268].
  %%CITATION = HEP-PH 0404268;%%

\bibitem{Maltoni:2004ei}
  M.~Maltoni, T.~Schwetz, M.~A.~Tortola and J.~W.~F.~Valle,
  %``Status of global fits to neutrino oscillations,''
  New J.\ Phys.\  {\bf 6} (2004) 122,
  for an update see arXiv:hep-ph/0405172 v5;
  %%CITATION = HEP-PH 0405172;%%
%
%\bibitem{Schwetz:2006dh}
  T.~Schwetz,
  %``Global fits to neutrino oscillation data,''
  Phys.\ Scripta {\bf T127}, 1 (2006)
  [arXiv:hep-ph/0606060].
  %%CITATION = HEP-PH 0606060;%%

\bibitem{Masina:2005hf}
  I.~Masina, in \cite{sumrule}.

\bibitem{Ahmed:2003kj}
  S.~N.~Ahmed {\it et al.}  [SNO Collaboration],
  %``Measurement of the total active B-8 solar neutrino flux at the Sudbury
  %Neutrino Observatory with enhanced neutral current sensitivity,''
  Phys.\ Rev.\ Lett.\  {\bf 92} (2004) 181301
  [arXiv:nucl-ex/0309004].
  %%CITATION = NUCL-EX 0309004;%%

\bibitem{Araki:2004mb}
  T.~Araki {\it et al.}  [KamLAND Collaboration],
  %``Measurement of neutrino oscillation with KamLAND: Evidence of spectral
  %distortion,''
  Phys.\ Rev.\ Lett.\  {\bf 94}, 081801 (2005)
  [arXiv:hep-ex/0406035].
  %%CITATION = HEP-EX 0406035;%%

\bibitem{Beacom:2003nk}
  J.~F.~Beacom and M.~R.~Vagins,
  %``GADZOOKS! Antineutrino spectroscopy with large water Cherenkov
  %detectors,''
  Phys.\ Rev.\ Lett.\  {\bf 93}, 171101 (2004)
  [arXiv:hep-ph/0309300].
  %%CITATION = HEP-PH 0309300;%%

\bibitem{Choubey:2004bf}
  S.~Choubey and S.~T.~Petcov,
  % ``Reactor anti-neutrino oscillations and gadolinium loaded  Super-Kamiokande
  %detector,''
  Phys.\ Lett.\ B {\bf 594} (2004) 333
  [arXiv:hep-ph/0404103].
  %%CITATION = HEP-PH 0404103;%%

\bibitem{Petcov:2006gy}
  S.~T.~Petcov and T.~Schwetz,
  %``Precision measurement of solar neutrino oscillation parameters by a
  %long-baseline reactor neutrino experiment in Europe,''
  Phys.\ Lett.\ B {\bf 642} (2006) 487
  [arXiv:hep-ph/0607155].
  %%CITATION = HEP-PH 0607155;%%
  
\bibitem{Minakata:2004jt}
  H.~Minakata, H.~Nunokawa, W.~J.~C.~Teves and R.~Zukanovich Funchal,
  %``Reactor measurement of theta(12): Principles, accuracies and physics
  %potentials,''
  Phys.\ Rev.\ D {\bf 71}, 013005 (2005)
  [arXiv:hep-ph/0407326].
  %%CITATION = HEP-PH 0407326;%%

\bibitem{Bandyopadhyay:2004cp}
  A.~Bandyopadhyay, S.~Choubey, S.~Goswami and S.~T.~Petcov,
  % ``High precision measurements of Theta(solar) in solar and reactor  neutrino
  %experiments,''
  Phys.\ Rev.\ D {\bf 72} (2005) 033013
  [arXiv:hep-ph/0410283].
  %%CITATION = HEP-PH 0410283;%%

\bibitem{Huber:2002mx}
  P.~Huber, M.~Lindner and W.~Winter,
  %``Superbeams versus neutrino factories,''
  Nucl.\ Phys.\ B {\bf 645} (2002) 3
  [arXiv:hep-ph/0204352].
  %%CITATION = HEP-PH 0204352;%%

%\cite{Huber:2004ka}
\bibitem{Huber:2004ka}
  P.~Huber, M.~Lindner and W.~Winter,
  % ``Simulation of long-baseline neutrino oscillation experiments with
  %GLoBES,''
  Comput.\ Phys.\ Commun.\  {\bf 167}, 195 (2005)
  [arXiv:hep-ph/0407333].
  %%CITATION = HEP-PH 0407333;%%

\bibitem{Campagne:2006yx}
  J.~E.~Campagne, M.~Maltoni, M.~Mezzetto and T.~Schwetz,
  %``Physics potential of the CERN-MEMPHYS neutrino oscillation project,''
  arXiv:hep-ph/0603172.
  %%CITATION = HEP-PH 0603172;%%

\bibitem{Barger:2006vy}
  V.~Barger {\it et al.}, 
  % M.~Dierckxsens, M.~Diwan, P.~Huber, C.~Lewis, D.~Marfatia and B.~Viren,
  %``Precision physics with a wide band super neutrino beam,''
  Phys.\ Rev.\ D {\bf 74}, 073004 (2006)
  [arXiv:hep-ph/0607177].
  %%CITATION = HEP-PH 0607177;%%

\bibitem{Diwan:2003bp}
  M.~V.~Diwan {\it et al.},
  %``Very long baseline neutrino oscillation experiments for precise
  %measurements of mixing parameters and CP violating effects,''
  Phys.\ Rev.\ D {\bf 68} (2003) 012002
  [arXiv:hep-ph/0303081].
  %%CITATION = HEP-PH 0303081;%%

\bibitem{Burguet-Castell:2005pa}
  J.~Burguet-Castell, D.~Casper, E.~Couce, J.~J.~Gomez-Cadenas and P.~Hernandez,
  %``Optimal beta-beam at the CERN-SPS,''
  Nucl.\ Phys.\ B {\bf 725}, 306 (2005)
  [arXiv:hep-ph/0503021].
  %%CITATION = HEP-PH 0503021;%%

\bibitem{Huber:2006wb}
  P.~Huber, M.~Lindner, M.~Rolinec and W.~Winter,
  %``Optimization of a neutrino factory oscillation experiment,''
  Phys.\ Rev.\ D {\bf 74} (2006) 073003
  [arXiv:hep-ph/0606119].
  %%CITATION = HEP-PH 0606119;%%

\bibitem{Cervera:2000vy}
 A.~Cervera, F.~Dydak and J.~Gomez Cadenas,
  %``A large magnetic detector for the neutrino factory,''
  Nucl.\ Instrum.\ Meth.\ A {\bf 451} (2000) 123.
  %%CITATION = NUIMA,A451,123;%%

\bibitem{Huber:2003ak}
  P.~Huber and W.~Winter,
  %``Neutrino factories and the 'magic' baseline,''
  Phys.\ Rev.\ D {\bf 68}, 037301 (2003)
  [arXiv:hep-ph/0301257].
  %%CITATION = HEP-PH 0301257;%%

\bibitem{Minakata:2001qm}
  H.~Minakata and H.~Nunokawa,
  %``Exploring neutrino mixing with low energy superbeams,''
  JHEP {\bf 0110} (2001) 001
  [arXiv:hep-ph/0108085].
  %%CITATION = HEP-PH 0108085;%%

\bibitem{Huber:2005ep}
  P.~Huber, M.~Maltoni and T.~Schwetz,
  %``Resolving parameter degeneracies in long-baseline experiments by
  %atmospheric neutrino data,''
  Phys.\ Rev.\ D {\bf 71}, 053006 (2005)
  [arXiv:hep-ph/0501037].
  %%CITATION = HEP-PH 0501037;%%
   
\end{thebibliography}
\end{document}